\documentclass[reprint,amsmath,amssymb,aps,longbibliography,showpacs]{revtex4-1}

\usepackage{graphicx}
\usepackage{dcolumn}
\usepackage{bm}
\usepackage{CJK}
\usepackage{subfigure}
\usepackage{color}
\usepackage{epstopdf}

\begin{document}

\preprint{APS/123-QED}
\title{Band-gap structure and chiral discrete solitons in optical lattices with artificial gauge fields}

\author{Qinzhou Ye$^{1,2}$}
\author{Xizhou Qin$^{1}$}
\author{Yongyao Li$^{3}$}
\author{Honghua Zhong$^{1}$}
\author{Yuri S. Kivshar$^{4}$}
\author{Chaohong Lee$^{1,2,4}$}
\altaffiliation{Email: lichaoh2@mail.sysu.edu.cn; chleecn@gmail.com}

\affiliation{$^{1}$TianQin Research Center \& School of Physics and Astronomy, Sun Yat-Sen University (Zhuhai Campus), Zhuhai 519082, China}
\affiliation{$^{2}$Key Laboratory of Optoelectronic Materials and Technologies, Sun Yat-Sen University (Guangzhou Campus), Guangzhou 510275, China}
\affiliation{$^{3}$Department of Applied Physics, South China Agricultural University, Guangzhou 510642, China}
\affiliation{$^{4}$Nonlinear Physics Centre, Research School of Physics and Engineering, Australian National University, Canberra ACT 2601, Australia}

\date{\today}

\begin{abstract}
  We study three-leg-ladder optical lattices loaded with repulsive atomic Bose-Einstein condensates and subjected to artificial gauge fields.
  By employing the plane-wave analysis and variational approach, we analyze the band-gap structure of the energy spectrum and reveal the exotic swallow-tail loop structures in the energy-level anti-crossing regions due to an interplay between the atom-atom interaction and artificial gauge field.
  Also, we discover stable discrete solitons residing in a semi-infinite gap above
  the highest band, these discrete solitons are associated with the chiral edge currents.
\end{abstract}

\maketitle
\section{Introduction}

It is of great interest in realizing gauge potentials for neutral atoms~\cite{Lin_Nature_462_628,Dalibard_RMP_83_1523,Lin_Nature_471_83,Wang_PRL_109_095301}, which may open a way to mimic the effects of magnetic fields acting on charged particles via neutral atomic systems. The gauge potentials bring nontrivial characters for single-particle dispersion and induce several novel quantum phenomena from single- to many-body levels. Also, the spin-orbital (SO) coupling may give rise to double-well dispersion structure and Dirac points, which lead to anisotropic critical superfluid velocity~\cite{Higbie_PRL_88_090401}, zitterbewegung effects~\cite{LeBlanc_NJP_15_7,Qu_PRA_88_021604} and semiclassical spin Hall effect~\cite{Beeler_Nat_498_201}.
Taking into account the atom-atom interactions in the SO coupled Bose-Einstein condensates (BECs), there appear several macroscopic quantum phenomena including vortices~\cite{Xu_PRL_107_200401,Radic_PRA_84_063604}, monopoles~\cite{Conduit_PRA_86_201605}, domains~\cite{Wang_PRL_105_160403}, skyrmions~\cite{Kawakami_PRL_109_015301}, and solitons~\cite{Achilleos_PRL_110_264101,Xu_PRA_87_013614}.

More recently, the artificial gauge fields in optical lattices have attracted extensive attentions. The effective magnetic fields in optical lattices have been achieved by laser-assisted tunneling~\cite{Jaksch_NJP_5_56,Aidelsburger_PRL_107_255301,Aidelsburger_PRL_111_185301,Miyake_PRL_111_185302} and dynamical shaking~\cite{Hauke_PRL_109_145301,Struck_Sci_333_996}.
In addition, the artificial gauge fields have also been created in optical superlattices~\cite{Atala_NP_10_588} and in optical lattices with synthetic dimension~\cite{Celi_PRL_112_043001,Mancini_Science_349_1510,Stuhl_Science_349_1514}.
These lattice systems with artificial gauge fields are well suitable for exploring the Hofstadter butterfly~\cite{Hofstadter_PRB_14_2239}, quantum Hall states~\cite{Thouless_PRL_49_405}, topological insulators~\cite{Hansan_RMP_82_3045} and topological phase transitions~\cite{Atala_NP_9_795,Jotzu_Nat_515_237,Aidelsburger_NP_11_162}.
In particular, it has been demonstrated that the gauge fields may induce exotic chiral currents~\cite{Atala_NP_10_588} and chiral edge states~\cite{Mancini_Science_349_1510,Stuhl_Science_349_1514}.

Due to the atom-atom interaction, these systems provide a promising platform to explore the interplay between gauge field, nonlinearity and lattice potential. It has been demonstrated that, due to the interplay between nonlinearity and lattice potential~\cite{Wu_PRA_64_061603,Trombettoni_PRL_86_2353}, exotic swallow-tail loops appear in the band-gap structures~\cite{Wu_PRA_65_025601,Diakonov_PRA_66_013604,Mueller_PRA_66_063603, Machholm_PRA_67_053613,Watanabe_Entropy_18_118}.
Moreover, due to the balance between nonlinearity and dispersion, spatially localized nonlinear modes can exist in the forbidden gap~\cite{Zobay_PRA_59_643,Louis_PRA_67_013602,Eiermann_PRL_92_230401}.
In the last three years, many interestingly macroscopic quantum phenomena of BECs in optical lattices subjected artificial gauge fields have been explored. Lots of stable gap and gap-stripe solitons in SO coupled BECs in spin-dependent optical lattices, which can be classified according to physical symmetries, have been found~\cite{Kartashov_PRL_111_060402,Lobanov_PRL_112_180403}. Based upon a generalized discrete nonlinear Schrodinger equation including various types of SO couplings, it has been demonstrated the existence of discrete solitons with different miscibilities~\cite{Belicev_JPB_48_065301} and symmetries~\cite{Salerno_PLA_379_2252}. However, there are still many important open questions, such as, (1) How the interplay between the gauge field and the nonlinearity affects the band-gap structure? and (2) Are there stable chiral discrete solitons in the energy gap?

In this paper, we consider atomic BECs in a three-leg ladder model with an artificial magnetic flux and study their band-gap structures and chiral discrete solitons. We find that, due to an interplay between the nonlinear atom-atom interaction and artificial gauge fields, swallow-tail loops appear in the band-gap structures. In addition, we numerically demonstrate the existence of stable discrete solitons in the semi-infinite gap above energy band. In a contrast to the solitons studied earlier, the discrete solitons in our system show chiral edge currents.
Moreover, different from the uniform chiral edge currents in a non-interacting system, the chiral edge currents in our system are spatially localized in the discrete solitons. Additionally, we analyze stability of the discrete solitons by employing the linear stability analysis.

The paper is organized as following. In Sec. II, we describe our model. Based upon the stationary extended states, we give the band-gap structures for different gauge fields and interaction strengths in Sec. III. In Sec. IV, we study stable discrete solitons and discuss their properties. Finally, we summarize our results in Sec. V.

\begin{figure}[htb]
\includegraphics[width=8.6 cm]{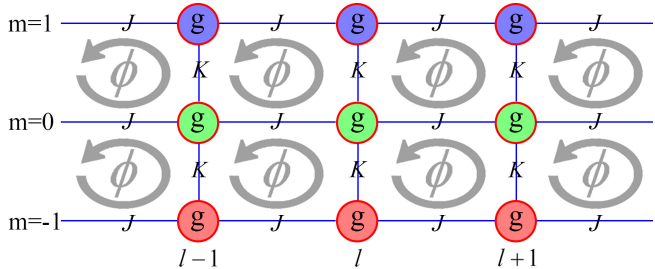}
\caption{\label{model} (Color online) Schematic diagram for a three-leg BEC ladder subjected to artificial gauge field with a flux of $\phi$ per plaquette. Here $J$ and $K$ denote respectively the intra- and inter-chain tunneling strengths. The colors red, green, blue stand for the atoms in the three chains respectively. The on-site interaction strength is represented by $g$.}
\end{figure}

\section{Model}

We consider atomic BECs held in a three-leg ladder subjected to an artificial uniform magnetic field, see Fig.~\ref{model}.
Under the tight-binding condition, the system obeys the following Hamiltonian:
\begin{equation}
\begin{split}
H=&-J\sum_{l=1}^{L_x}\sum_{m=-1}^{1}(e^{-im\phi}\hat{b}^{\dagger}_{l+1,m}\hat{b}_{l,m}+H.c.)\\&-K\sum_{l=1}^{L_x}\sum_{m=0}^{1}(\hat{b}^{\dagger}_{l,m-1}\hat{b}_{l,m}+H.c.)\\
&+\frac{g}{2}\sum_{l=1}^{L_x}\sum_{m=-1}^{1}\hat{n}_{l,m}(\hat{n}_{l,m}-1).
\label{Hamiltonian}
\end{split}
\end{equation}
Here, $\hat{b}_{l,m}$ annihilates a boson on the site $(l,m)$, $\hat{n}_{l,m}=\hat{b}^\dagger_{l,m}\hat{b}_{l,m}$ is the particle number operator,
the on-site interaction strength $g$ is determined by the s-wave scattering length,
$J\exp(-im\phi)$ and $K$ are respectively the intra- and inter-chain hopping strengths,
and $\phi$ is the artificial magnetic flux which can be created by using laser-assisted tunneling~\cite{Atala_NP_10_588} .
Obviously, when a particle hops around a plaquette, the accumulated phases sum up to $\phi$, which provides an uniform effective magnetic field to the system.
Without loss of generality, we consider each leg has $L_x$ sites and label the three legs with $m=\{-1,0,1\}$.
Experimentally, the inter-chain hopping strength $K$ can be manipulated by adjusting the intensity of the lasers which create the optical lattice, the intra-chain hopping strength $J$ and the flux $\phi$ can be tuned by adjusting the amplitude and angle of the Raman lasers respectively, and the on-site interaction strength $g$ can be controlled by utilizing the Feshbach resonance~\cite{Chin_RMP_82_1225}.

For the weakly interacting systems, the above Hamiltonian can be analyzed by the mean-field (MF) approach.
In the MF treatment, one can adopt the Glauber's coherent states as the variational trial states, that is,
\begin{equation}
\begin{split}
&|\Phi\rangle=\bigotimes_{l,m}|\Psi_{l,m}\rangle,\\
&|\Psi_{l,m}\rangle=e^{-\frac{1}{2}\left|\Psi_{l,m}\right|^2}\sum_{n=0}^{\infty}(\frac{\Psi_{l,m}^n}{\sqrt{n!}} |n\rangle).
\end{split}
\end{equation}
Here, $\Psi_{l,m}=\langle\Phi|\hat{b}_{l,m}|\Phi\rangle$ and $\Psi^*_{l,m}=\langle\Phi|\hat{b}^{\dagger}_{l,m}|\Phi\rangle$ are the order parameters for the site $(l,m)$.
Therefore the system obeys the MF Hamiltonian,
\begin{equation}
\begin{split}
H=&-J\sum_{l=1}^{L_x}\sum_{m=-1}^{1}(e^{-im\phi}\Psi^*_{l+1,m}\Psi_{l,m}+H.c.)\\
&-K\sum_{l=1}^{L_x}\sum_{m=0}^{1}(\Psi^*_{l,m-1}\Psi_{l,m}+H.c.)\\
&+\frac{g}{2}\sum_{l=1}^{L_x}\sum_{m=-1}^{1}\left|\Psi_{l,m}\right|^4.
\label{Hamiltonian_MF}
\end{split}
\end{equation}
Assume the total particle number is $N=\sum_{l,m} \left|\Psi_{l,m}\right|^2$, the average particle number per site is given as $n=N/3L_x$.
Below, for simplicity, we adopt the parameter $J$ as a unit to rescale the other parameters.
Obviously, for the evolution time, it will be rescaled in unit of $\omega_J^{-1}=\hbar/J$.

\section{Band-gap Structure}

Here, we show how the band-gap structure is affected by the interplay between the nonlinearity and gauge fields.
To give the band-gap structure, one has to impose the periodic boundary condition on the chain direction, which ensures the translational invariance along that direction.
For our nonlinear lattice system, there may appear two typical kinds of stationary states: the extended states in form of plane waves which preserve the symmetry of the Hamiltonian, and the spatially localized state such as discrete solitons which do not preserve the symmetry of the Hamiltonian.
The band-gap structure is determined by the extended stationary states.

\begin{figure*}[htbp]
\includegraphics[width=2.0\columnwidth]{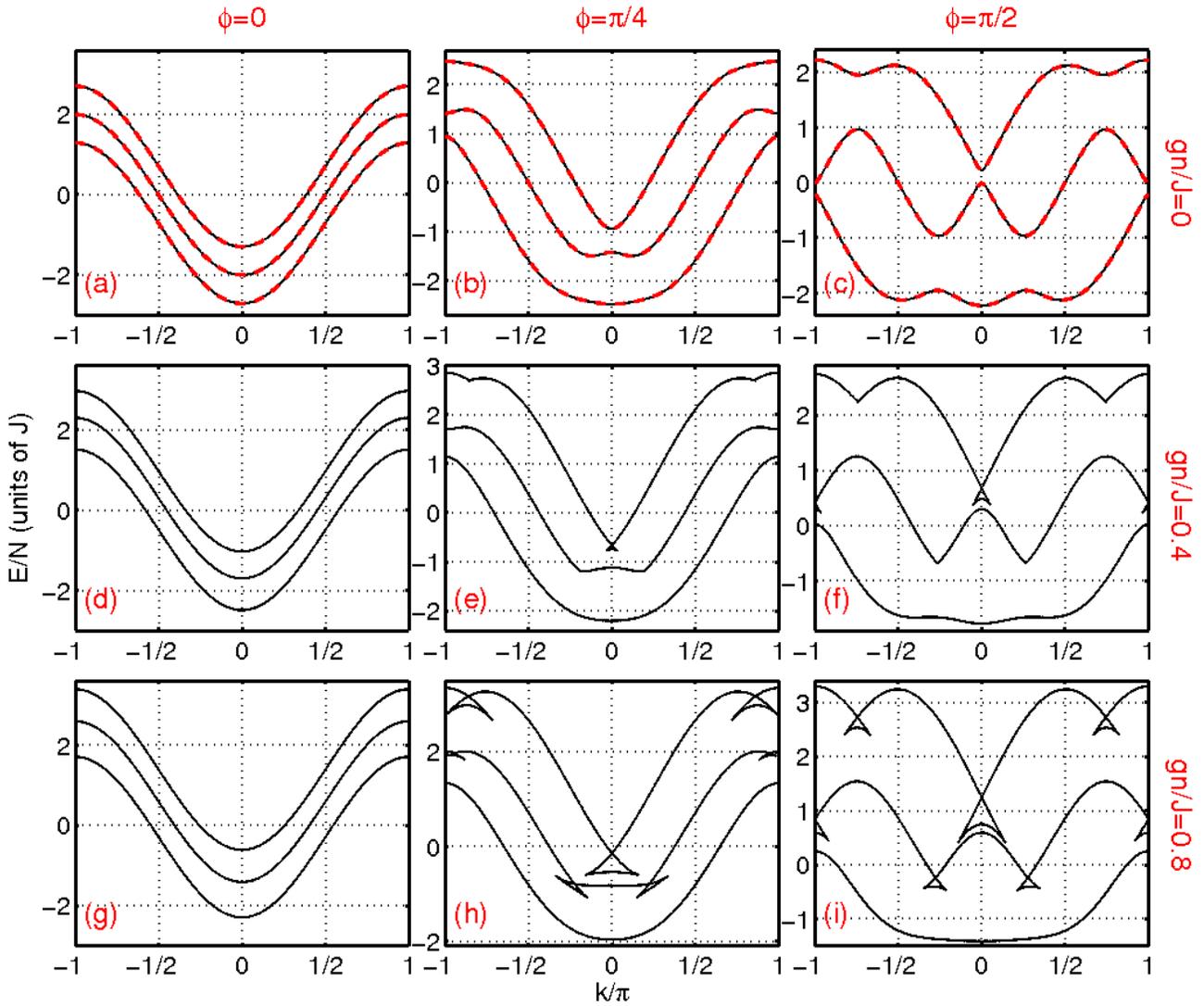}
  \caption{\label{spectrum_combine} (Color online) Band-gap structures for different interaction strengths and magnetic fluxes.
  The three columns correspond to the magnetic flux $\phi=0$,$\pi/4$ and $\pi/2$, respectively.
  The three rows correspond to the interaction strength $gn/J=0$, $0.4$ and $0.8$, respectively.
  (a)-(c) show the dispersion relations for the non-interacting system (i.e. $g=0$), in which the dashed red and black solid lines respectively denote the analytical and variational results.
  Other parameters are chosen as: $K/J=0.5$, $N=10000$.}
\end{figure*}

In the absence of interaction, the dispersion relation can be directly given by diagonalize the second quantized Hamiltonian~(\ref{Hamiltonian}).
Attribute to the translational invariance, the quasi-momentum $k$ along the chain direction is a good quantum number and its first Brillouin zone can be given as $[-\pi, \pi]$.
Based upon the Fourier transform,
\begin{equation}
\begin{split}
\hat{b}_{l,m}=\frac{1}{\sqrt{L_x}}\sum_k e^{ikl}\hat{a}_{k,m},~~k=2\pi n/L_x,\\
n=\left(-\frac{L_x}{2},-\frac{L_x}{2}+1,...,\frac{L_x}{2}-1\right).
\label{FT}
\end{split}
\end{equation}
the Hamiltonian~(\ref{Hamiltonian}) can be written as
\begin{equation}
\begin{split}
\hat{H}=\sum_k\sum_m[&-K(\hat{a}_{k,m}^{\dagger}\hat{a}_{k,m+1}+H.c.)\\
&-2J\hat{a}_{k,m}^{\dagger}\hat{a}_{k,m}\cos(k+m\phi)].
\end{split}
\end{equation}
As the quasi-momentum $k$ is a good quantum number, the corresponding Hamiltonian for a specific quasi-momentum $k$ reads as
\begin{equation}
\begin{split}
\hat{H}_k=\sum_m[&-K(\hat{a}_{k,m}^{\dagger}\hat{a}_{k,m+1}+H.c.)\\
&-2J\hat{a}_{k,m}^{\dagger}\hat{a}_{k,m}\cos(k+m\phi)].
\end{split}
\end{equation}
In the basis $\{|m\rangle=\hat{a}_{k,m}^{\dagger}|0\rangle\}$ (where $|0\rangle$ is the vacuum state), the matrix elements for $\hat{H}_k$ are given as
\begin{equation}
\begin{split}
\langle m_1|\hat{H}_k|m_2\rangle=&-K(\delta_{m_1+1,m_2}+\delta_{m_1,m_2+1})\\
&-2J\delta_{m_1,m_2}\cos(k+m_1\phi).
\end{split}
\end{equation}
The dispersion relation (band-gap structure) can be obtained by calculating the eigenvalues of $\hat{H}_k$.
In Fig.~\ref{spectrum_combine}~(a), (b) and (c), given $J$ and $K$, we show the dispersion relation (red dashed lines) for different fluxes $\phi$.
The energy spectrum contains three bands, in which the energy gaps are determined by the inter-chain tunneling strength $K$.
The gauge field shifts the band minimum from the zero momentum states to nonzero momentum states.
Fixing the hopping strengths $J$ and $K$, when the gauge field $\phi$ increases, the lowest band changes from a single-well to a triple-well structure and the second band changes from a single-well to a double-well structure.
This dispersion resembles the one for SO coupled spin-1 BECs~\cite{Lan_PRA_89_023630}.

Taking into account the atom-atom interaction, although the quasi-momentum $k$ is still a good quantum number, it is difficult to directly diagonalize the eigen-equations for each $k$.
However, one can obtain the dispersion relation by using the variational method, which request to give an initial guess of the stationary states and then solve the variational equations by extremizing the energy function.
For a given quasi-momentum $k$, we introduce the variational ansatz as,
\begin{equation}
\Psi_{l,m}^k=\sqrt{\frac{N}{L_x}}a_{m}^k e^{ikl}.
\label{ansatz}
\end{equation}
Here, $k=2\pi n/L_x$ with $n=\left(-\frac{L_x}{2},-\frac{L_x}{2}+1,...,\frac{L_x}{2}-1\right)$ and $a_m^k$ are complex amplitudes satisfying the normalization condition,
\begin{equation}
\sum_{m=-1}^{+1}\left|a_{m}^k\right|^2=1.
\label{normalization}
\end{equation}
Inserting the variational wave function~(\ref{ansatz}) into the MF Hamiltonian~(\ref{Hamiltonian_MF}), the energy function is given as,
\begin{equation}
\begin{split}
E(k) = &-2JN\sum_{m}\left|a_m^k\right|^2cos(k+m\phi)\\
&-KN\sum_{m}[(a_{m-1}^k)^*a_m^k+c.c.]
+\frac{gN^2}{2L_x}\sum_{m}\left|a_m^k\right|^4.
\label{energy_functional}
\end{split}
\end{equation}
One can use the Lagrange multiplier method to extremize the energy function~(\ref{energy_functional}) under the normalization condition~(\ref{normalization}).
By introducing a new variable $\lambda$, we construct the Lagrange function,
\begin{equation}
I=\mathcal{E}-\lambda(1-\sum_{m}\left|a_m^k\right|^2),
\end{equation}
with the energy per atom $\mathcal{E}=E/N$.
For a stationary state, the derivation of the Lagrange function to the variational variables $(\lambda,\{a_m^k\})$ should be zero.
Therefore, we can get the variational equations,
\begin{equation}
\begin{split}
-2Ja_{m}^k\cos(k+m\phi)-K(a_{m-1}^k+a_{m+1}^k)&\\
+3gn\left|a_{m}^k\right|^2a_{m}^k+\lambda a_{m}^k&=0,\\
\label{Variational_equ}
\end{split}
\end{equation}
for a given quasi-momentum $k$.
Combining the variational equations with the normalization condition, one can determine the values of $a_m^k$ and then give the dispersion relation.

In Fig.~\ref{spectrum_combine}, we show the band-gap structures for different interaction strengths $gn/J$ and different magnetic fluxes $\phi$.
For noninteracting systems, the analytical band-gap structures can be obtained by exactly diagonalizing the Hamiltonian matrix~(7).
When the magnetic flux $\phi$ increases, the lowest band gradually changes from a single-well shape to a three-well one and the second band gradually changes from a single-well shape to a double-well one.
To check the validity of our variational method, we also numerically calculate the corresponding band-gap structures by using our variational method.
We find that our numerical results are well consistent with the analytical ones, see the top row of Fig.~\ref{spectrum_combine}.

Then we calculate the band-gap structures for interacting systems.
Without loss of generality, we focus on discussing the repulsive interaction.
It is known that, for atomic BECs in one-dimensional optical lattices without gauge fields, the atom-atom interaction may result the appearance of swallow-tail loops at the avoided crossing points~\cite{Wu_PRA_65_025601,Diakonov_PRA_66_013604, Mueller_PRA_66_063603, Machholm_PRA_67_053613}.
Here, for atomic BECs in our three-leg ladder, there is no swallow-tail loop structure if the gauge field is absent (see the left column of Fig.~\ref{spectrum_combine}).
However, if the gauge fields are introduced, swallow-tail loop structures appear due to the cooperation between the gauge fields and the atom-atom interactions [see panels (e), (f), (h) and (i) of Fig.~\ref{spectrum_combine}], in which the per-particle interaction energy $(gn)$ is competitive with respect to the per-particle hopping energy $(J, K)$.
The interaction energy can influence the distribution of the atoms and at the same time itself also depends on the local atomic number, so that, the corresponding stationary states will be very different from the noninteracting case and there may appear extra stationary solutions which induce the swallow-tail loop structures.
Given nonzero magnetic fluxes: $\phi=\pi/4$ (middle column) and $\phi=\pi/2$ (right column), in the vicinities of the level crossings, the upper bands become sharper, the low bands become flatter and swallow-tail loop structures appear in upper bands, when the interaction strength increases from $gn/J=0$ (first row) to 0.4 (second row) and then 0.8 (third row).
It should be pointed out that the attractive interaction may also bring swallow-tail loop structures at the avoided crossings, however, those swallow-tail loop structures appear in lower bands instead of upper bands.

The positions of the swallow-tail loop structures are determined by the magnetic flux $\phi$.
To have a better understanding, we consider the case of no inter-chain tunneling $(K=0)$ at first.
Thus the three chains are decoupled and all three chains have sinusoidal dispersion relations whose minima are shifted in different directions.
There are crossings between three dispersion curves and the crossing positions are determined by $\phi$.
If $\phi=0$, the three dispersion curves completely overlap each other.
If $\phi\neq0$, the three dispersion curves cross at the points where $k=0, \pm\pi, \pm\phi/2$ and $\pm(\pi-\phi/2)$.
Nonzero inter-chain tunneling $(K\ne0)$ will open these crossings and sufficiently strong interaction will induce swallow-tail loop structures in these opened crossings.
Different from the multi-band systems of BECs in one-dimensional optical lattices without gauge field, whose swallow-tail loop structures appear at the center and/or boundary of the Brillouin zone, the swallow-tail loop structures in our system may appear at arbitrary positions in the Brillouin zone by adjusting the magnetic flux $\phi$.

\section{Chiral Discrete Solitons}

In addition to the swallow-tail loop structures, which can be understood as a kind of macroscopic quantum self-trapping caused by the nonlinearity, spatially localized in-gap modes may appear as a result of the balance between the dispersion and the nonlinearity.
For BECs in an optical lattice, the hoppings between different lattice sites play the role of dispersion and the atom-atom interaction brings the nonlinearity and stationary gap-solitons have been observed in experiments~\cite{Eiermann_PRL_92_230401}.
For a deep optical lattice, by expanding the wave-function with the Wannier states, one can get a discrete nonlinear Sch\"{o}rdinger equation and the localized states in such discrete system are called discrete solitons.
On the other hand, the gauge fields may bring novel chiral behaviors~\cite{Atala_NP_10_588,Mancini_Science_349_1510,Stuhl_Science_349_1514}.
It is still unknown that how the gauge fields affect the spatially localized modes of BECs in an optical lattice?
Below, we will study the spatially localized modes in our three-leg BEC ladder subjected gauge fields and explore the existence of stable chiral discrete solitons.

\begin{figure}[htbp]
\includegraphics[width=8.6 cm]{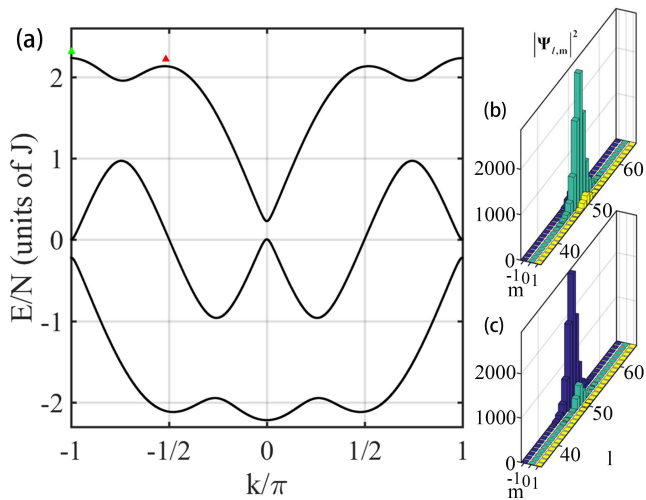}
\caption{\label{gap_soliton_combine} (Color online) Band-gap structure and the density profiles of discrete gap-solitons. (a) Band-gap structure and positions of discrete gap-solitons. (b) Density profile for the discrete soliton localized in the bulk chain $(m=0)$, which is labelled by the green triangle in (a). (c) Density profile for the discrete soliton localized in the edge chain $(m=-1)$, which is labelled by the red triangle in (a). The parameters are chosen as: $N=10000$, $K/J=0.5$, $\phi=0.5\pi$ and $gn/J=0.006$.}
\end{figure}

From the MF Hamiltonian~(\ref{Hamiltonian_MF}), we have the complex amplitudes  $\Psi_{l,m}$ obey the time-evolution equations,
\begin{equation}
\begin{split}
i\frac{\mathrm{d}\Psi_{l,m}}{\mathrm{d}t}=&-J(e^{-im\phi}\Psi_{l-1,m}+e^{im\phi}\Psi_{l+1,m})\\
&-K(\Psi_{l,m-1}+\Psi_{l,m+1})+g\left|\Psi_{l,m}\right|^2\Psi_{l,m}.
\label{evolution_equ}
\end{split}
\end{equation}
It is difficult to get analytical soliton solutions for the above nonlinear discrete Schr\"{o}dinger equations.
Therefore one has to employ numerical technique to find soliton solutions, such as, the Newton-conjugate-gradient method~\cite{Yang_JCP_228_7007}, the modified squared-operator method~\cite{Yang_SAM_118_153} and the imaginary-time propagation method.
In our calculation, we choose the parameters as $\phi=\pi/2$ and $gn/J=0.006$.
We find that discrete solitons localize in both chain and rung direction, with their centers in either the bulk chain ($m=0$) or the edge chains ($m=\pm1$), as shown in Fig.~\ref{gap_soliton_combine}(b) and (c).
The density profile of the discrete solitons have a bell shape along the chain direction and can be well fitted by a $\text{sech}^2$ function.
The soliton width decreases quickly as the interaction strength increases and finally all atoms will stay in a single site for sufficiently strong interaction.

Because of the repulsive interaction, the discrete solitons all exist in the semi-infinite gap above the highest band.
As a result of the gauge field, our numerical results show that the soliton phase linearly changes along chain direction.
This means that the discrete solitons can be described by
\begin{eqnarray}
\Psi_{l,m}=f_{l,m}e^{ikl},
\end{eqnarray}
in which $f_{l,m}$ are real amplitudes and $e^{ikl}$ describe the phase distribution.
The linear phase distribution indicates that each discrete soliton has a certain quasi-momentum and energy.
In Fig.~\ref{gap_soliton_combine}(b) and (c), we show the density profiles of two discrete solitons, whose quasi-momenta and energies are respectively labelled by the green and red triangles in the band-gap structure shown in Fig.~\ref{gap_soliton_combine}(a).
Obviously, the discrete solitons exist in the semi-infinite gap above the maximums of the highest band.
In comparison with the extended states for the corresponding quasi-momenta, the discrete solitons have similar population distribution in the rung direction but are spatially localized in the chain direction.

Now we discuss how the particle currents of the discrete solitons depend on the gauge field.
In a non-interacting three-leg ladder, chiral edge currents have been observed in the quench process~\cite{Mancini_Science_349_1510,Stuhl_Science_349_1514}.
Similarly, for our MF system, one can define the currents as
\begin{eqnarray}
\begin{split}
{j}_{l,l+1;m} = i J(\Psi_{l+1,m}^{*}\Psi_{l,m}e^{-i\phi m}-c.c.),\\
{j}_{l;m,m+1} = i K(\Psi_{l,m+1}^{*}\Psi_{l,m}-c.c.),
\end{split}
\end{eqnarray}
where ${j}_{l,l+1;m}$ stands for the currents along chain direction and ${j}_{l;m,m+1}$ is the currents along the rung direction.
The average currents in each chain read as
\begin{equation}
j_m=\frac{1}{L_x}\sum_l {j}_{l,l+1;m}.
\end{equation}
and the net chiral current is given as, $j_C=j_{1}-j_{-1}$.

\begin{figure}[htbp]
\includegraphics[width=8.6 cm]{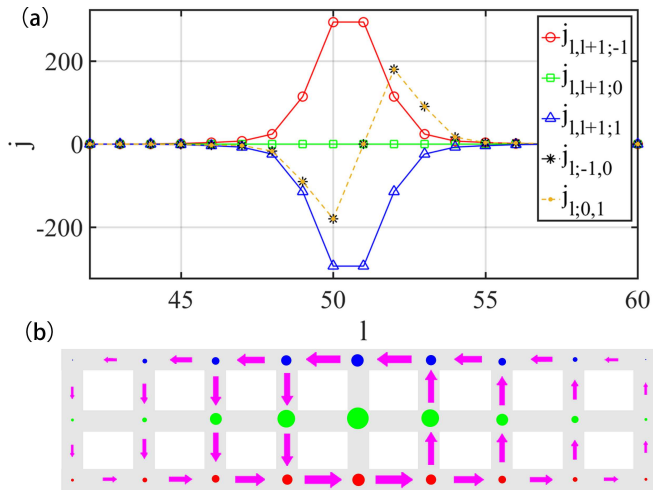}
\caption{\label{currents} (Color online) Currents of the discrete soliton shown in Fig.~\ref{gap_soliton_combine}(b). (a) The currents along the chain and rung direction of the discrete soliton. (b) The schematic of the discrete soliton and its currents. The area of each circle qualitatively represents the atomic number for each lattice site. The direction and length of the arrows respectively denote the direction and magnitude of the currents.}
\end{figure}

Through calculating the currents between all neighboring lattice sites, we find that chiral edge currents are still preserved in the discrete solitons, although they are spatially localized.
However, different from the extended states, whose currents are uniform in the chain direction, the currents of discrete solitons are spatially localized currents along the chain direction.
In Fig.~\ref{currents}, we plot the current distribution of the discrete soliton shown in Fig.~\ref{gap_soliton_combine}(b).
These currents show similar chirality as the non-interacting systems, that is, the currents in two edge chains $(m = \pm 1)$ always flow in opposite directions and there is almost no currents in the bulk chain $(m=0)$.

The current directions of discrete solitons depend on the gauge field.
When the magnetic flux $\phi$ changes its sign, the currents will reverse their directions, which confirm such currents have chirality.
Remarkably, since the currents is not uniform along the chain direction any more, there are nonzero currents along the rung direction to balance the currents flowing in and out each site.
Therefore, we call these discrete solitons of chiral currents as chiral discrete solitons.

Since stable solutions are usually easier to be generated and observed in experiments, it is crucial to analyze the stability of the chiral discrete solitons we have got.
Below we will employ the linear stability analysis to analyze the stability of our chiral discrete solitons under small perturbations.
We write the perturbed state in the form of
\begin{eqnarray}
\Psi_{l,m}(t)=e^{-i\mu t}(\psi_{l,m}+u_{l,m}e^{\lambda t}+v_{l,m}^*e^{\lambda ^*t}),
\end{eqnarray}
with $\mu$ denoting the chemical potential.
Here, the complex amplitudes $|u_{l,m}|$ and $|v_{l,m}|$ are far smaller than the stationary state $\psi_{l,m}$.
Substituting the above perturbed state into the time-evolution equations~(\ref{evolution_equ}), one can linearize the equations around the stationary state by reserving the linear terms of $u_{l,m}$ and $v_{l,m}$.
Comparing the coefficients for the terms of $e^{\lambda t}$ and $e^{\lambda^*t}$, one can obtain the eigen-equation,
\begin{eqnarray}
\begin{split}
&(-2ig\left|\psi_{l,m}\right|^2+i\mu)u_{l,m}\\
&~~+iJe^{-im\phi}u_{l-1,m}+iJe^{im\phi}u_{l+1,m}\\
&~~+i K u_{l,m-1}+i K u_{l,m+1}-i g \psi_{l,m}^2 v_{l,m}=\lambda u_{l,m},\\
&(2ig\left|\psi_{l,m}\right|^2-i\mu)v_{l,m}\\
&~~-iJe^{im\phi}v_{l-1,m}-iJe^{-im\phi}v_{l+1,m}\\
&~~-i K v_{l,m-1}-i K v_{l,m+1}+i g (\psi_{l,m}^2)^* v_{l,m}=\lambda v_{l,m}.
\label{stability_eigen_equation}
\end{split}
\end{eqnarray}
According to the linear stability criteria, the stationary state is stable if all eigenvalues $\{\lambda\}$ are purely imaginary, otherwise the stationary state is unstable if one or more eigenvalues have nonzero real parts.

\begin{figure}[htbp]
\includegraphics[width=8.6 cm]{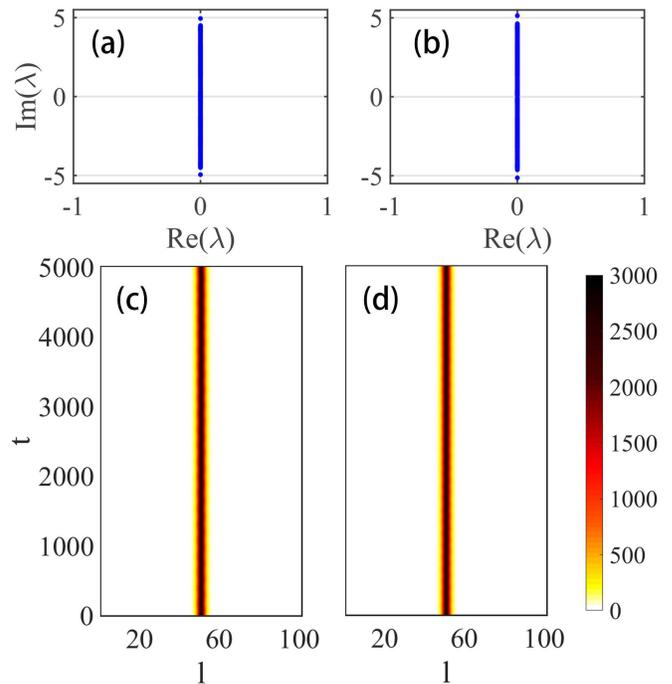}
\caption{\label{stability} (Color online) The linear-stability spectrum and the long-time evolution of the perturbed discrete solitons. (a) and (b): the linear-stability spectrum (blue points) for the two discrete solitons shown in Fig.~\ref{gap_soliton_combine}(b) and (c). (c) and (d): the long-time evolutions of the perturbed discrete solitons. Where, for simplicity, we only show the time-evolutions for the chain of most atoms in each discrete soliton. The color bar stands for the atomic number.}
\end{figure}

For the two discrete solitons shown in Fig.~\ref{gap_soliton_combine}(b) and (c), we calculate all eigenvalues $\{\lambda\}$ by solving the eigen-equation~(\ref{stability_eigen_equation}).
The eigenvalues for these two discrete solitons are shown in Fig.~\ref{stability}(a) and (b), respectively.
We find that all eigenvalues are purely imaginary numbers, which indicates that these two discrete solitons are linearly stable.

To check the validity of the linear stability analysis, one can examine the stability of a stationary state via the long-time evolution of the perturbed stationary state.
By imposing $1\%$ random-noise perturbations onto the two discrete solitons shown in Fig.~\ref{gap_soliton_combine}(b) and (c), we simulate their long-time evolutions, see Fig.~\ref{stability}(c) and (d).
Our numerical results show that the density profiles are almost unchanged.
This indicates that these discrete solitons are stable under the small perturbations, which consist with the predictions of the linear stability analysis. In our simulation, we have chosen the running time as 5000 dimensionless time units.
Considering the recent experiment~\cite{Atala_NP_10_588}, the laser assisted tunneling amplitude $J$ along the chain direction is about $66$Hz, which means that 5000 dimensionless time units is about 10s.
Such a long time is sufficiently long for manipulating and detecting these chiral discrete solitons.

\section{Concluding remarks}

We have studied the stationary macroscopic quantum states of atomic BECs in a three-leg ladder subjected to artificial magnetic fields.
We have obtained the band-gap structure for the extended states and explore the existence of the swallow-tail loop structures.
Different from the previous swallow-tail loop structures in the center or boundary of the Brillouin zone, our loop structures may appear at arbitrary positions determined by the gauge fields. Moreover, we find that stable discrete solitons may appear in the semi-infinite gap over the highest energy band and these discrete solitons have chiral edge currents due to the presence of the gauge fields.

Based upon the current available techniques for ultracold atom manipulations, we believe that it would be possible to observe these novel chiral discrete solitons. By combining the techniques of optical lattices with the light modulated potential \cite{Islam_Nature_528_77}, the three-leg ladder can be realized, by using laser-assisted tunneling~\cite{Jaksch_NJP_5_56,Aidelsburger_PRL_107_255301,Aidelsburger_PRL_111_185301,Miyake_PRL_111_185302} and dynamical shaking~\cite{Hauke_PRL_109_145301,Struck_Sci_333_996}, for creating the desired gauge fields.
The swallow-tail loop structures can be demonstrated by observing the adiabatic destruction of the Bloch oscillation, and the localized chiral edge currents can be detected through the quench dynamics~\cite{Atala_NP_10_588}. Furthermore, an $N$-leg ladder can also be realized by internal hyperfine states of alkaline-earth(-like) atoms~\cite{Celi_PRL_112_043001}, in which the interaction have $SU(N)$ invariance.
Also, it is interesting to explore how the gauge fields and $SU(N)$-invariant interaction affect the band-gap structure and spatially localized modes.

\begin{acknowledgments}

The authors thanks Dr Elena Ostrovskaya for helpful comments. This work was supported by the National Basic Research Program of China (Grant No. 2012CB821305), the National Natural Science Foundation of China (Grants No. 11374375, No. 11574405, No. 11465008 and No. 11575063), and the Australian Research Council.

\end{acknowledgments}


%

\end{document}